# Statistical Properties of a Polymer Globule Formed during Collapse with the Irreversible Coalescence of Units

# Statistical Properties of a Polymer Globule Formed during Collapse with the Irreversible Coalescence of Units

A. M. Astakhov[a,b], S. K. Nechaev[c,d,\*], and K. E. Polovnikov[b,e]

[a] *Semenov Institute of Chemical Physics, Russian Academy of Sciences, Moscow, 119991 Russia*
[b] *Faculty of Physics, Moscow State University, Moscow, 119991 Russia*
[c] *Independent University of Moscow, Moscow, 119002 Russia*
[d] *Lebedev Physical Institute, Russian Academy of Sciences, Moscow, 119991 Russia*
[e] *Skolkovo Institute of Science and Technology, Moscow, 143026 Russia*
\*e-mail: sergei.nechaev@gmail.com

**Abstract**—Collapse of the polymer chain upon the sharp decrease of solvent quality is studied. During collapse, any pair of polymer units appearing in a sufficiently close vicinity in space has the possibility with a certain probability to form an irreversible crosslink, thereby preventing the interpenetration of chain material between the forming clusters. Globular structures having different spatial chain packing at various scales are obtained by computer simulations. It is shown that the dependence of probability of contact between two monomers in space $P(s)$, where $s$ is a distance between monomers along chain, reproduces a number of characteristic features observed previously in experiments on the analysis of three-dimensional chromatin packing. The cluster analysis of intramolecular contact maps makes it possible to express the hypothesis that there are characteristic discrete hierarchical levels in polymer packing associated with the number-theoretic origin of rare-event statistics and inherent to individual maps of intra- and interchromosomal contacts.

## INTRODUCTION

The statistical physics of macromolecular compounds in the Soviet Union was born three times. For the first time, this was in 1959 after publication of M.V. Volkenstein's book *Configurational Statistics of Polymer Chains* [1], for the second time in 1964 following publication of the monograph *Conformations of Macromolecules* by T.M. Birshtein and O.B. Ptitsyn [2], and finally for the third time in 1968 together with I.M. Lifshitz's seminal paper "Some Problems of Statistical Theory of Biopolymers" [3].

The statistical physics of macromolecules emerged from the need for predicting diverse physicochemical properties of polymeric materials. Just the second birth in 1964 defined the characteristic features of the field which having been developed for almost six decades at the onset of the 2000s at last began to answer questions posed upon its appearance, namely, made it possible to design tailor-made materials. There would be nothing surprising in this natural development of science area, if not for one circumstance: classical studies which laid the basis of polymer physics and modern studies of sensitive issues concerning the design of specific macromolecular structures with desired properties are the works of the same person—Tatiana Maksimovna Birshtein.

One of science themes in which Tatiana Maksimovna has been engaged for many years and which has passed as a leitmotif through most of her studies (from the investigation of configurational statistics of synthetic polymers to the research of biopolymers with specified physicochemical properties) is the collapse/decollapse of polymer chains of different architecture (both polyelectrolyte brushes [4] and linear chains under action of the applied force [5]).

This paper addresses specific features of the long-standing collapse of the homopolymer chain containing a certain amount of functional groups able to form bonds. Namely, after the sharp decrease of solvent quality in the consecutive formation of the globular phase monomers appearing to be in close proximity in space with a certain probability may coalesce (form an irreversible "crosslink"). As a result, the collapse gives rise to a growing stable network of bonds which prevents the spatial interpenetration between parts of a macromolecule. The typical final structure of the polymer prepared under these conditions, even at not high length, has a well-defined hierarchical pattern. This makes it possible to conjecture about the biological relevancy of the irreversible crosslinking mechanism in the collapse process during formation of topologically associating domains in chromatin. Below, we will discuss in detail the consecutive collapse process, examine the behavior of different statistical properties of final globular structures, and compare the obtained results with some mechanisms behind formation of the so-called chromatin "territories" in DNA packing which are proposed in a number of modern biological studies. To the best of our knowledge, no data are available on the systematic analysis of similar model systems from formation of the folded hierarchical spatial structure of a macromolecule to the study of fractal structure of clusters in contact maps.

## LINEAR CHAIN COLLAPSE WITH CONSECUTIVE IRREVERSIBLE CROSSLINKING

The crumpled (fractal) globular state of a macromolecule resembling the three-dimensional statistical Peano curve or the Hilbert curve [6] with fractal dimension $D_f = 3$ in three-dimensional space was proposed in [7] as an equilibrium structure of the unknotted polymer ring in a poor solvent. Despite many testing attempts in real and computer experiments, the crumpled globule existed for almost two decades (from 1988 to 2007) mostly as a purely mathematical hypothesis. The interest of biologists in the fractal structure was aroused in 1993 by paper [8]. According to the authors of [8], the packing of DNA chain in chromosome may resemble a crumpled globule. This collapsed state consists of hierarchical set of crumples and is thermodynamically equilibrium for a ring chain and kinetically frozen for a linear chain, in which various segments of a macromolecule are weakly entangled in a broad scale range. The hypothesis of hierarchical packing has provided an answer to one of the principal issues related to copying of information from DNA during transcription.

The attitude towards the crumpled globule hypothesis changed substantially since publication of paper [9] in 2009, in which, on the basis of analysis of population contact maps of human chromatin obtained experimentally by the Hi–C method (genome-wide chromosome conformation capture method) [10], it was found that average scaling probability $P(s)$ for a contact inside DNA decays approximately as $P(s) \sim s^{-1}$, where $s$ is the genomic distance (i.e., the distance along chain) between nucleotides (groups of nucleotides). In accordance with [9], generally, in three-dimensional space, for chain packing with the fractal dimension $D_f$, the dependence $P(s)$ should take the form $P(s) \sim s^{-3/D_f}$. Thus, the statistical analysis of genome maps made it possible to suggest that chromatin is packed in chromosome as a crumpled (fractal) globule with dimension $D_f = 3$. This study caused a rebirth of interest to the crumpled globule and its various modifications in the context of chromatin folding and led to the formulation of a number of constructive ideas concerning the effective interactions stabilizing the certain fractal dimension of the packing [11, 12].

### Main Features of Crumpled Globule and Its Place in Modern Genomics

Let us recall basic ideas related to formation of the crumpled structure of unknotted ring macromolecule in a confined volume or in a poor solvent. Let us consider a closed unknotted non-self-intersecting polymer chain composed of $N$ monomers. In a poor solvent ($B < 0$), there exists a certain critical chain length $g^*$ depending on temperature and the energy of volume interactions such that chains with lengths above $g^*$ collapse. As was shown in [7], if we take a sufficiently long chain, then $g^*$-mer units may be considered as new block monomer units (crumples of the minimal scale). Let us consider a part of a chain containing several block monomers. This new part of the chain should again collapse "in itself"; that is, it should form a crumple of the next scale if other parts of the chain do not embed into a given part. A chain of new subblocks (crumples of the new scale) collapses again and again until the chain as a whole forms a crumple of the largest scale into which all initial monomers are collected. The line which represents the chain trajectory is constructed according to the above scheme (Fig. 1a). This is the three-dimensional statistical analog of the well-known self-similar Peano curve. Figure 1b schematically shows the landscape in which conformational transitions between various groups of monomers take place. For example, in order to add a group of monomers 7 and 8 (as a whole) to the forming globule, the entropy barrier $w_2$ should be overcome. The most important specific feature of this landscape is its hierarchical structure. This hierarchical packing makes itself evident as the fact that, at all scales beginning from $g^*$ and up to the maximal scale, chain segments are unentangled and densely fill volume assigned to them. It could be assumed that spatial fluctuations destroy this thermodynamically unfavorable scale-invariant structure; however, as was shown in [7], if the number of units in the crumple is above $N_e$, then crumples do not mix but remain segregated in space. For various models and systems, the typical values of phenomenological parameter $N_e$ having the sense of the average number of units between entanglements neighboring along chain are in the range of 30–300.

In real and computer experiments, parameter $N_e$ is high; therefore, the reliable visualization of the hierarchical crumpled structure at different scales requires polymers with a considerable length, with the maximum number of units on the order of $N_{max} \sim 10^5$ and the dimensionless density on the order of 0.5. Because of a considerable computational complexity for such long chains, the Monte Carlo study of statistical and correlation properties in the knotted ring polymer in the globular phase was first performed only in 2015 [13], in which on the basis of analysis of the behavior of radius of gyration of chain segment in the globule the estimated value of $N_e \sim 300$ was obtained.

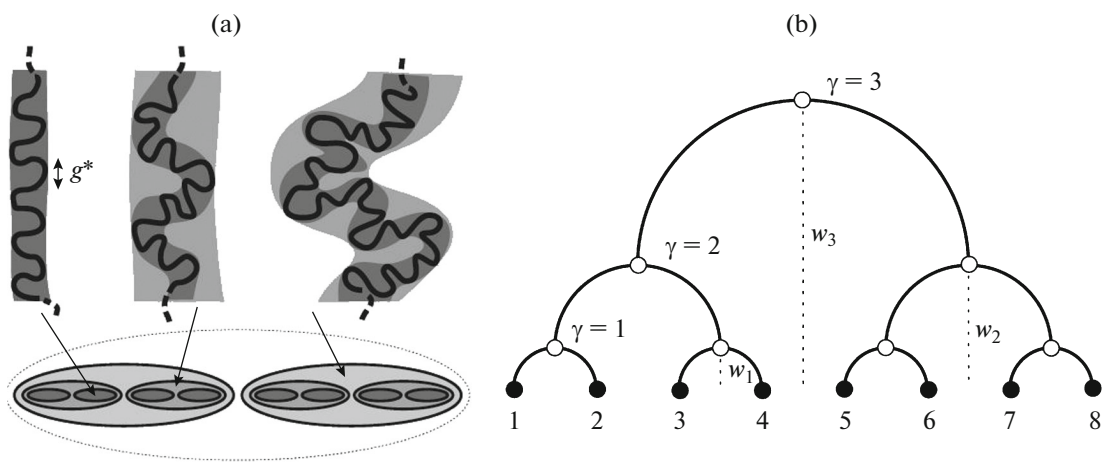

**Fig. 1.** (a) Consecutive steps in formation of the crumpled globule and (b) the hierarchical structure of entropy barriers between folds of different scales.

For the linear open-ended polymer, the folded state is not equilibrium; that is, under a sharp decrease of temperature below the θ point, the folded structure is formed as a metastable "kinetically frozen" state which emerges at times of order of $N^2$ and then relaxes to the ordinary equilibrium globule via reptations in times of order of $(N/N_e)^3$.

Linear polymers (both rings and open-ended chains) containing the maximal number of units $N_{max} \lesssim 10 N_e$ do not form a reliable hierarchical crumpled structure. This is largely due to the fact that the clusters (crumples) formed on a chain aggregate and mix; as a result, the hierarchical crumpled structure is not preserved and the collapse of polymer chains resembles the gradual growth and coalescence of beads-crumples in the bead-on-the-spring model of the polymer. Growth occurs until all pearls coalesce into a single large droplet absorbing all other droplets. The collapse of a sufficiently long chain was studied in detail in computer simulations, for example, in [14]. In accordance with [14], under sharp worsening of solvent quality, the conventional collapse of the polymer chain gives rise to the metastable state which lacks the crumpled spatial structure but which is still practically unknotted. In [15], the stable fractal globular states of chains were obtained using a special potential of volume interactions with the dependence of force acting between monomers on distance along the chain.

Concluding this short (and far from being complete) review of possible equilibrium and metastable states of linear and ring polymers obtained during the collapse process, let us note two important points.

(1) Only long unknotted rings $N_{max} \geq 300 N_e$ form the thermodynamically reliable crumpled hierarchical state.

(2) Both the linear chains and unknotted rings of the intermediate length $N_{max} \lesssim 10 N_e$ during the collapse process form drops sequentially coalescencing with time which are ordinary globules without the hierarchical structure.

Although the crumpled globule is our favorite child, it should honestly be noted that it does not explain all details of chromatin packing and certainly its place among modern concepts consistent with new experimental observations in genomics should be found. The theoretical models of chromatin packing in a nucleus which provide an explanation for the observed behavior of Hi-C intrachromosomal contact maps are conditionally divided into two groups. The first group of studies [16–22] relies on specific interactions within chromatin, such as formation of loops or bridges, and the authors of these publications do not appeal to the crumpled globule concept. Another group of researchers tends to explain the structure of chromatin in terms of large-scale topological interactions [8, 13, 23, 28] within the framework of the crumpled polymer globule approach. In order to demonstrate that the hierarchical folding of the crumpled globule is consistent with the fine structure of experimentally observed Hi–C maps, we combined the assumption that chromatin may be regarded as a heteropolymer with the frozen primary sequence and the general mechanism of hierarchical polymer folding [29]. Using this hypothesis, we managed to describe large-scale compartmentalization of chromosomes which was not expected a priori.

An almost three-decade history of the crumpled globule may be summed up by the classical phrase: "The king died. Long live the king!" Using the crumpled globule concept, a long journey took place from the abstract nonperturbative description of topological interactions in globular polymers to the study of specific issues of chromatin folding. Along this path, the relation between the statistics of polymer constraints and Brownian bridges in the non-Euclidean geometry was identified, new results for probability distributions of random braids were derived, and many features of DNA packing in chromosomes were explained. At the same time, the emergence of new experimental data clearly shows that our initial topological arguments were too rough and naive. However, they gave birth to new understanding of the role of topology in genomics and led to the appearance of new ideas and methods which underlie the basis of modern statistical topology of polymers.

### Territorial Organization of Chromatin and Model Description

Recent studies [30, 31] make it possible to state that the description of chromatin in the cell nucleus as a homogeneous dense globule is strongly simplified: in reality, there exist both unfolded and dense chromatin regions referred to as topologically associating domains (TADs). The latest investigations indicate that there is a well-defined relationship between the type of packing and the information value of the genetic material. For example, for *Drosophila*, a meaningful correlation between the transcriptional activity of chromosome regions and their accessible "loose" packing for annotated population Hi-C contact maps was observed experimentally [32]. On the other hand, inactive chromatin forms TADs. However, it should be admitted that finer details of DNA packing inside TADs and ideas about mechanisms leading to the predominant collapse of inactive chromatin remain at a level of hypotheses so far.

One the most realistic concepts of chromatin folding is the hypothesis that its three-dimensional packing structure is associated not only with the large length of chains but also with specific interactions between monomer units, specifically topological crosslinks which are formed with participation of a protein complex—cohesin—and whose presence is confirmed experimentally [33]. The detailed mechanism of their functioning is not yet fully understood; as one of the plausible hypotheses, it is described by the loop extrusion model [35]. According to this model, crosslinks existing for a finite time may be conditionally represented as small rings which translocate along chromosome, so that the size of the loop that is pulled through the ring increases until the breakdown of the crosslink occurs or it encounters any outer stop factor (e.g., the "collision" of rings takes place). In other words, the model assumes that crosslink play the role of molecular nanomotors and their mechanism is reduced to the extrusion ("pulling through") of loops. There are various modifications of the loop extrusion model. For example, in [36], which agrees well with the experimental data, it is stated that, under specific conditions of the new crosslinks formation, the size of loops pulled through the ring increases via the simple diffusion of rings as well.

There are other models in which the collapse of the polymer chain was studied in the presence of additional interactions—crosslinks. In the context of chromatin folding, the model of the randomly crosslinked Rouse chain was considered analytically [37]. Within the framework of this model, the degree of chain crosslinking can be estimated using the experimental data on the probability of contact between two monomers remote along chain. However, note that the Rouse model disregards the excluded volume of chain which strongly affects polymer packing in the globular phase. For the Rouse phantom chain, a computer model of the irreversible collapse via crosslinking was also constructed which in many respects is similar to our model [38]. We must also mention paper [39], in which the model of coalescence of drops (blobs) was considered as applied to the polymer chain collapse and it was shown that this approach is applicable to the coarse-grained collapse simulation. Of particular note is study [34], in which the model of chain collapse with a small number of preformed crosslinks and chain pulling by ends was used to describe the behavior of dense domains of inactive chromatin (TADs). The results obtained using this model describe well the experimental data on the structure of TADs in chromatin. Later we will demonstrate that our model yields a similar result without any stretching force.

In order to explore the process of collapse for sufficiently long chains with crosslinking, the method of Brownian dynamics without explicit solvent was used. Pairwise noncovalent interactions were set by the Lennard-Jones potential with $\sigma = 1$. Chemical bonds between neighboring monomers (and the newly formed crosslinks) were simulated by the harmonic potential with $k_{\text{spring}} = 200$ and $L_0 = 0.9$. The integration step was taken standard for similar models, $\Delta t = 0.005\tau$. The computer experimental protocol consisted of two main steps which will be described below.

(1) The initial state of the system was prepared; for this purpose, polymer chains generated as random walk without self-crossings were equilibrated for some time in the coil state in the good solvent regime ($\epsilon_{\text{LJ}}/kT = 0.1$); afterwards, the quality of the solvent worsened abruptly by increasing $\epsilon_{\text{LJ}}/kT$ to 1.

(2) Formation of crosslinks during the collapse process was performed as follows: at every $\tau$ steps of the simulation, all pairs of monomers occurring at a distance smaller than $R_{\text{cut}} = 2$ from each other were considered and a bond with probability $P$ was formed between these monomer pairs.

If probability $P$ was a time-independent constant and the number of contacts increased with time, the effective intensity of crosslinking would also increase with time. Eventually, this would lead to the formation of "frozen" conformations. But we were interested in sufficiently "soft" structures able to from a globule close to spherical at the end of collapse. Therefore, probability $P$ was selected to be sufficiently small and it decreased during collapse. A variation of $P(t)$ with time $t$ was selected so that the time dependence of the number of crosslinks would be practically linear (Fig. 2a) and the average number of new crosslinks formed during the entire collapse time would be on the order of 10% of the initial number of bonds (chain length). Chains with lengths of $10^4$ and $10^5$ units were considered; the crosslinking regime (dependence $P(t)$) was the same in both cases. Simulations were performed using the LAMMPS software package [40].

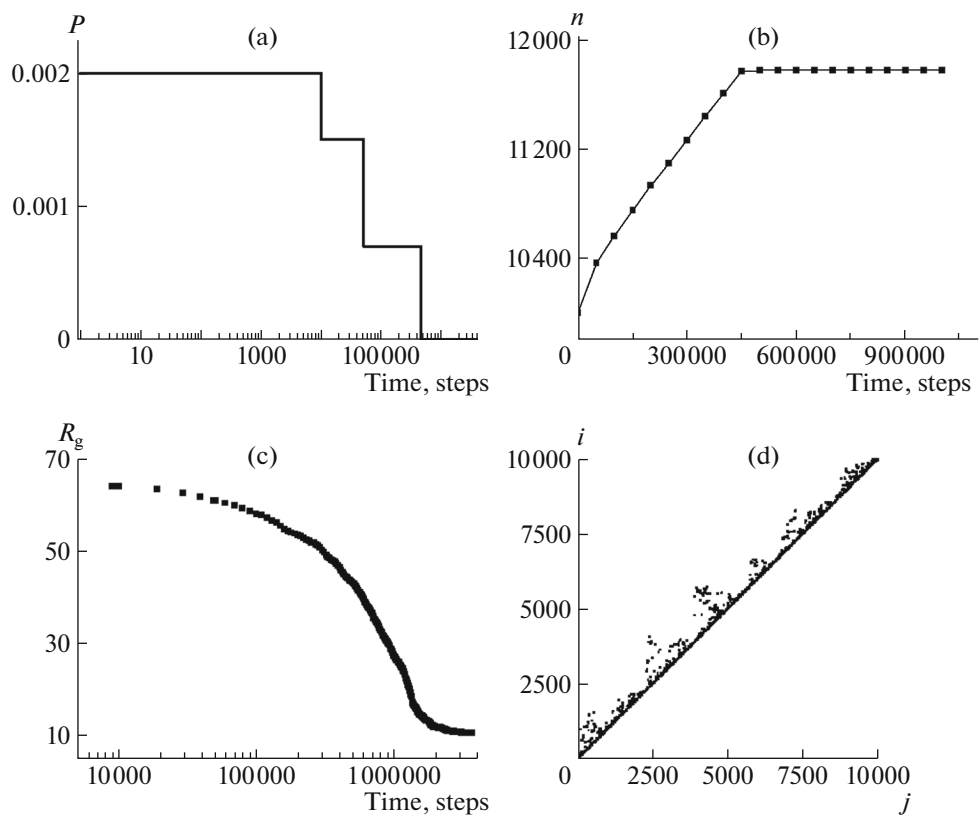

**Fig. 2.** Time dependences of (a) $P$ and (b) full number of bonds $n$ and examples of time dependence of (c) radius of gyration for $N = 10^4$ and (d) matrix of bonds (crosslinks) for $N = 10^4$.

For chains consisting of $N = 10^4$ units, computer experiments on collapse with consecutive crosslinking were performed. The results were compared with the data obtained for collapse without crosslinking. In both cases, the data were averaged over 20 independent realizations with different initial conformations. The characteristic time dependence of the radius of gyration for one of the realizations is presented in Fig. 2c. It is seen that the chain collapse proceeds at times on the order of $2 \times 10^6$ steps. For further analysis, conformations obtained at times on the order of $5 \times 10^6$ steps after solvent quality worsening were taken. The crosslink matrix for one of realizations for the chain with length of $10^4$ units is shown in Fig. 2b. A greater part of crosslinks is situated close to the principal diagonal (i.e., is formed between monomers situated at a sufficiently short distance along chain). This is due to the fact that the most intense formation of crosslinks is observed at the initial stage of collapse. As follows from Fig. 2b, the density of crosslinks is heterogeneous along chain, and regions which were already dense structures (blobs) at the previous stages of crosslinking appear to be more crosslinked.

## RESULTS AND DISCUSSION

### Polymer Collapse with the Irreversible Coalescence of Units

We compared the dependences of the radius of gyration of a subchain $R_g(s)$ and the probability of monomer contact $P_{cont}(s)$, where $s$ is the distance between monomers along chain, for the events of collapse with irreversible crosslinking and without it (Fig. 3). All dependences $R_g(s)$ show the Gaussian region of small length subchains, where $R_g(s) \sim s^{1/2}$, typical of the equilibrium globule; however, for longer subchains, a substantially different behavior is observed.

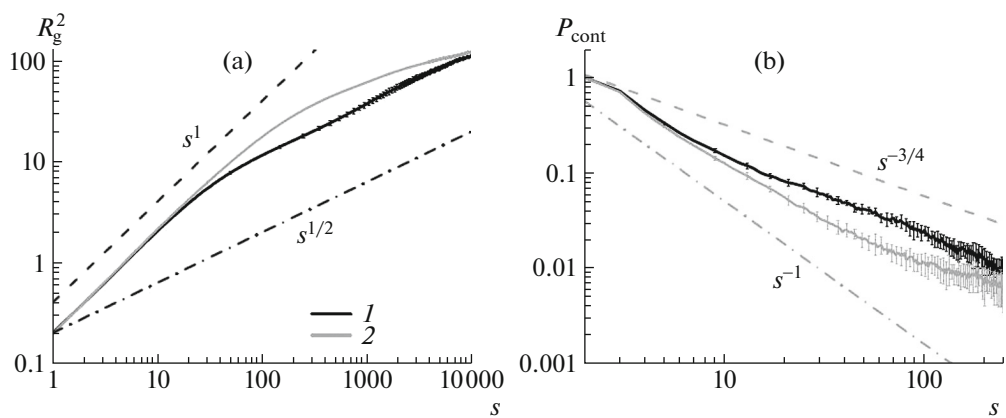

**Fig. 3.** Dependences of (a) the squared radius of gyration of subchain and (b) probability of contact on distance between monomers along chain $s$ in logarithmic coordinates for chain with $N = 10^4$. (*1*) Collapse with crosslinking and (*2*) collapse without crosslinking.

In the case of conventional collapse without crosslinking, there is a smooth saturation of dependence $R_g(s)$ related to the globule surface just the same as in many other experiments on the collapse of long chains [14].

For collapse with crosslinking for large loops, there is the dependence

$$R_g \sim s^{1/4} \quad (1)$$

The structure formed by the consecutive crosslinking of monomers appearing in close proximity in space during the collapse process after a sharp reduction in temperature below the θ point is schematically shown in Fig. 4. It is obvious that at large times monomers situated at a longer distance along chain are involved in crosslinking. In fact, the collapse process with crosslinking is a timing hierarchical coalescence typical of any renormalization procedure. As was shown in [41], dependence $R_g(s) \sim s^{1/4}$ is valid for ring polymers without volume interactions which form unentangled (contracting to a point) loops in the lattice of topological obstacles [43, 44]. The melts of ring polymers, in which each chain is in the lattice of obstacles created by other chains, were simulated in [42]. By forming crosslinks, we fixed the chain topology predominantly in the form of a treelike structure (Fig. 4). Note that, in the hierarchical collapse, the dependence $R_g \sim s^{1/4}$ makes itself evident only at a large length of chains: in long loops, the topology is fixed by crosslinks; however, these loops may rather easily interpenetrate because of a fairly insignificant role of volume interactions at these scales.

The dependence of contact probability on distance along chain (Fig. 3b) in the event of collapse with crosslinking at a fairly small length of subchains obeys the following pattern:

$$P_{cont} \sim s^{3/4} \quad (2)$$

This dependence was obtained in experiments on determining the structure of dense chromatin domains (TADs) and in the tension globule computer model [34], in which it was supposed that a chain with a small constant number of crosslinks collapses in the presence of the stretching force applied to polymer chain ends. Comparing this result with our data, it is natural to assume that, for scaling (1) to be realized, simple crosslinking of chain is quite enough (possibly with a special time-dependent protocol). Both scaling parameters, 1/4 in dependence (1) and 3/4 in dependence (2), correspond to an object with the fractal dimension $D_f = 4$. Certainly, this behavior in three-dimensional space is available only in a limited scale interval. To gain a more detailed insight into excluded volume effects, we also considered a chain with length of $N = 10^5$ units. Simulation was performed using the same parameters of crosslinking as for $N = 10^4$, and the results were averaged over five independent realizations after $8 \times 10^7$ simulation steps. It is clear that, in this case (Fig. 5), the dependence $R_g \sim s^{1/3}$ characteristic of the fractal globule shows itself within a broad range of scales. Note that estimation of $N_e$ in terms of this model is a very difficult problem, because this parameter depends on crosslink density and varies during the collapse process. In the final state, the system resembles the melt of crosslinked rings. On the basis of the plot shown in Fig. 5a, the value of $N_e$ may be estimated from the characteristic scale at which the chain statistics changes from Gaussian to folded; that is, $N_e \sim 30$. Note that, in the melt of unentangled rings, $N_e \sim 50$.

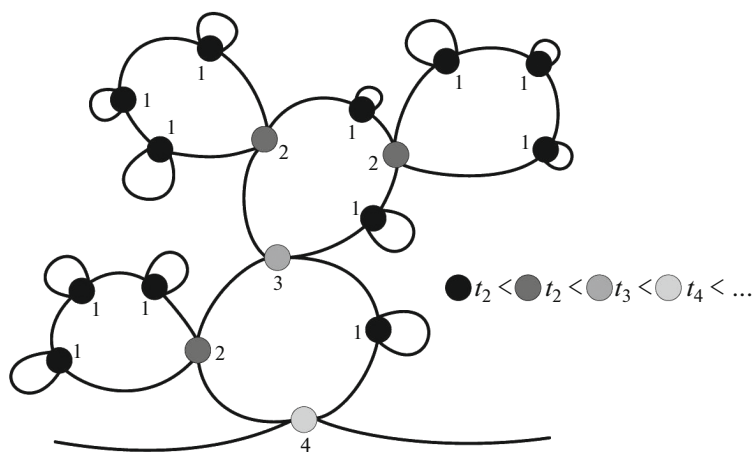

**Fig. 4.** Treelike structure formed via the consecutive crosslinking of monomers appearing to be side by side in space during the collapse process. Gray tints indicate crosslinks formed at different times.

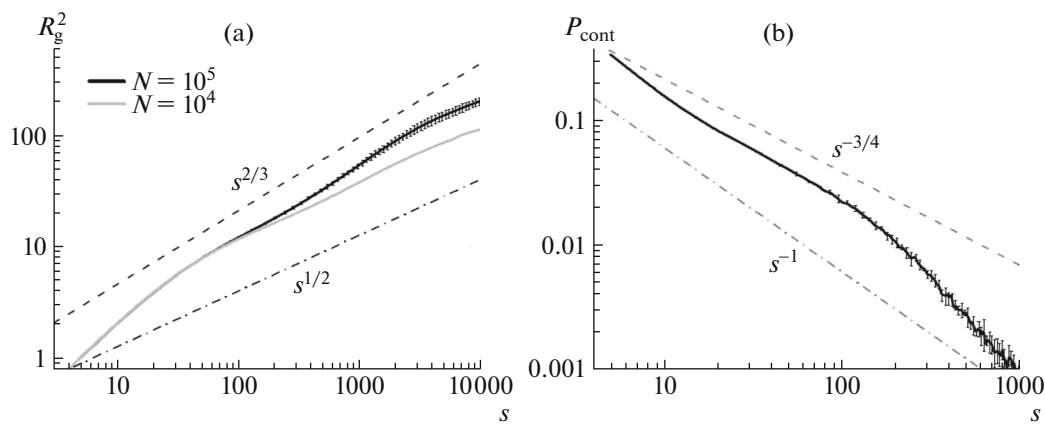

**Fig. 5.** Dependences of (a) the squared radius of gyration of subchain and (b) the probability of contact on distance between monomers along chain in logarithmic coordinates for chain with $N = 10^5$.

To explain the results shown in Fig. 5 let us estimate the subchain length, $s_0$, such that, in a chain divided into regions with length $s_0$, almost half of them will be "inside" the globule and the other half will have access to the surface (under assumption that all regions are globules). A rough estimate of $s_0$ may be derived from the relation $N^{2/3} = (N/s_0)s_0^{2/3}/4$. We believe that all subglobules are spherical and equate the total surface of the globule to the half surface of half subglobules. For chain $N = 10^5$, these considerations make it possible to evaluate the inflection point on the $R_g$ plot (chain access to the surface) $s_0 \approx 1500$. Naturally, at scales smaller than $s_0$ and larger than $g^*$, at which the chain always behaves as a Gaussian coil, blobs in a sufficiently crosslinked chain will not mix and the scaling of type $R_g \sim s^{1/4}$ will be impossible because a considerable part of subchains is surrounded by the polymer.

*Isolation of Cluster Structure in Contact Matrices*

To compare polymer configurations formed during collapses with irreversible crosslinking and without it, we analyzed the clusterization of contacts near the principal diagonal of maps analogous to "individual Hi-C maps." These maps became popular in recent years in the study of the three-dimensional organization of chromatin, because they provided complete information as to what monomer pairs are in contact in this configuration. By definition, the contact map with a certain configuration is matrix $A$, for which $A_{ij} = 1$ stands at the intersection of row $i$ and column $j$ if monomers $i$ and $j$ are in contact and $A_{ij} = 0$ in the

S32 ASTAKHOV et al.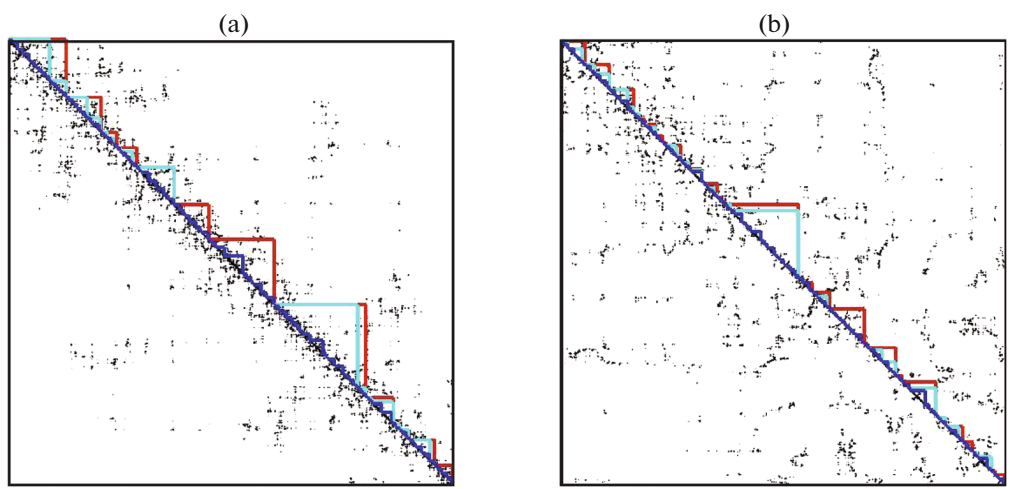

**Fig. 6.** Clusterization of contacts for $N = 10^4$ in (a) collapse with crosslinking and (b) collapse without crosslinking. In all cases, the initial region of chain length $N = 5 \times 10^3$ is shown. Three colors correspond to different levels of cluster nesting: red $r = 2/3$, blue-green $r = 5/7$, and blue $r = 3/4$. The higher the value of $r$, the higher the density of contacts inside clusters and the higher the level of hierarchy.

opposite case. In this study, we assume that both monomers are in contact if the distance between them is smaller than the cutoff radius $R_{cut} = 2\sigma$.

Clusters on the contact map were determined using the procedure of isolating series of sequences of regions (subclusters) near the diagonal with different critical values of contact density $r$. The procedure is as follows. At the first stage, we isolated all these monomer sequences $x_t, x_{t+1}, x_{t+2}, ..., x_{t+s-1}$ with length $s$ along chain for which the density of contacts inside the isolated region

$$\rho_t^{(s)} = \frac{1}{s^2} \sum_{i=t}^{s} \sum_{j=t}^{s} A_{ij} \quad (3)$$

was higher than the expected value of density for the sequence of a given size

$$\rho^{(s)} = \frac{1}{N-s} \sum_{i=1}^{N-s} \rho_i^{(s)} \quad (4)$$

and, moreover, it is higher than a certain critical value $r$ ($r < 1$) specifying the series number (the nesting level of clusters: denser clusters are nested into looser ones). Indeed, these sequences overlap strongly, especially for maps with a good clusterization. At the second stage, we unite the discovered subclusters into isolated domains using the OR operation: if a given monomer $x_i$ belongs to at least one of the "primary" sequences $\rho_t^{(s)}, t \leq i < t+s$, then $\mathbf{f}^{(r)}(x_i) = 1$; otherwise $\mathbf{f}^{(r)}(x_i) = 0$. Thus, vector $\mathbf{f}^{(r)}$ sets chain division into clusters at a given level of critical density $r$. Note that the exact boundaries of clusters $\{l_j\}$ and $\{r_j\}$ are determined by conditions $\mathbf{f}^{(r)}(l_j) = \mathbf{f}^{(r)}(r_j) = 0$ and, for any $k$, condition $l_j < k < r_j$ is fulfilled and $\mathbf{f}^{(r)}(k) = 1$.

Figure 6 presents the annotated contact maps for chains consisting of $N = 10^4$ monomers with and without crosslinking. Maps corresponding to collapse with crosslinking are characterized by a more defined linear structure. Moreover, the contacts of sequentially crosslinked chains clusterize much more strongly near the principal diagonal. These features result from the artificial limitation of the number of degrees of freedom of a chain: dynamically more probable contacts occur inside dense clusters that have already formed at the previous step. Thus, the consecutive crosslinking imposes formation of the hierarchical structure of clusters with different nesting levels. On the other hand, for collapse without crosslinking, contacts in each configuration are formed at all scales and do not lead to such a strong structuring.

Qualitative differences in map divisions are illustrated by dependences of the average cluster size $\langle L_r \rangle$ and the number of clusters $N_r$ on density parameter $r$ (Fig. 7). At small values of critical density ($r < 2/3$), only one large cluster with size of system $N$ may be distinguished. This is associated with a high average density of maps: sequences of the "initial hierarchy level" $\{x_i\}_t^{t+s-1}$ at small $r$ overlap strongly and unite into a common cluster. It may be said that the value of $r = 2/3$ corresponds to the percolation threshold in the sense of overlapping of sequences and the critical density of formation of an infinite cluster of contacts for maps of a given density. Upon further increase in

POLYMER SCIENCE, SERIES C Vol. 60 Suppl. 1 2018

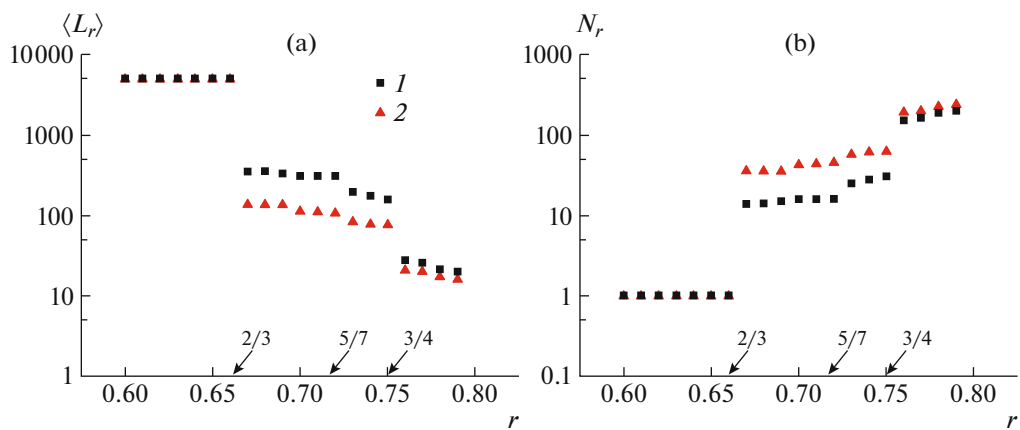

**Fig. 7.** (Color online) Dependences of (a) average cluster size $\langle L_r \rangle$ and (b) full number of annotated clusters $N_r$ as functions of density parameter $r$ for chain with length $N = 10^4$ (error is smaller than the point size). The values are presented for (*1*) collapse with crosslinking and (*2*) collapse without crosslinking.

critical density, the chain is discontinuously broken into clusters.

As follows from Fig. 7, at $2/3 < r < 3/4$ in the case of collapse without crosslinking, the formed clusters are approximately two times smaller, while their number is greater than that for collapse with crosslinking. In fact, in the latter case because of the hierarchical mechanism of crosslinking, contacts are clusterized to a more pronounced extent. This leads to a high density (above $r$ for each nesting level) of contacts inside large clusters. As was noted above, in collapse without additional fixation of crosslinks, contacts are formed less regularly (at all scales) and, as a consequence, the cluster structure turns out to be smeared out.

Upon further increase in density parameter $r > 3/4$, we observed another sharp transition as for the average size of clusters and their number. Namely, the average cluster size $\langle L_r \rangle$ for collapse with crosslinking becomes comparable with the fluctuation size corresponding to collapse without crosslinking. In other words, clusterization of the crosslinked polymer at small scales $s_{cr} \sim 10$ is similar to clusterization of a chain without crosslinking. The same effect may be observed on dependences of the average radius of gyration for the crosslinked polymer at both short $N = 10^4$ (Fig. 3a) and long chains $N = 10^5$ (Fig. 5a). Actually, at sufficiently small scales, the processes of excluded volume shielding prevail and the chain with hierarchical crosslinks does not feel cluster organization imposed on it. Therefore, in both cases, chain segments follow the Brownian statistics, $\langle R_g^2 \rangle \sim s$. It is natural to relate length $s_{cr}$ to the characteristic length of entanglement $s_{cr} \sim N_e$. At smaller scales, the chain is insensitive to topological constraints and, as a consequence, three-dimensional conformations of length segments $< s_{cr}$ for crosslinking and classical collapse events become indistinguishable.

It is interesting that the proposed method of chain division into clusters straightforwardly demonstrates the universal character of hierarchical relations in terms of "ultrametrics," that is, the hierarchy of nested clusters. By construction, clusters are ranked in terms of density: looser clusters contain denser clusters corresponding to the next hierarchy level. Upon going to the next level of cluster nesting, that is, during the breakup of clusters into finer ones, natural "washing out" of subclusters with a certain critical density $r_{cr}$ corresponding to transition occurs and the remaining sequences, whose integration determines chain division into clusters, have density a priori above $r_{cr}$. It turns out that the sequence of transition values $r_{cr} = \{2/3, 3/4, 4/5, \ldots\}$ is universal for various maps, as evidenced by the stepwise behavior of dependences shown in Fig. 7. The presence of this universal discrete set of transition values $r_{cr}^{(i)}$ is related to the number-theoretic features in the tail of distribution of contact densities inside size sequences from 2 to maximal $L$ near the diagonal (Fig. 8a). The appearing fractal structure is one of the incarnations of the so-called Thomae or "popcorn" function [47] $g(x)$, which is also known as the Riemann function, the raindrop function, the countable cloud function, the modified Dirichlet function, and the ruler function. This is one of the simplest number-theoretic functions having a nontrivial fractal structure (another known example is the everywhere continuous but not differentiable Weierstrass function). The Thomae function is deter-

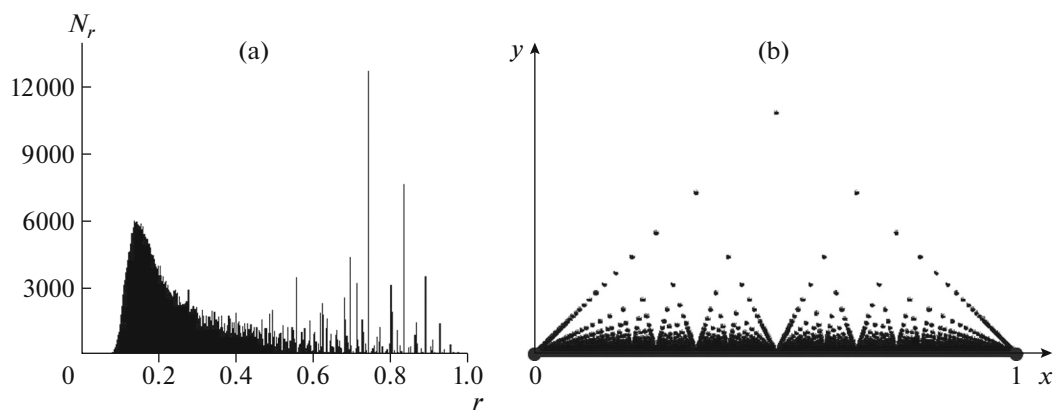

**Fig. 8.** (a) Dependence of the number of subclusters $N_r$ on density $r$ and (b) the Thomae function (5).

mined on the open interval $x \in (0,1)$ according to the following rule:

$$g(x) = \begin{cases} \frac{1}{q} & \text{if } x = \frac{p}{q} \text{ and } (p,q) \text{ mutually coprime} \\ 0 & \text{if } x \text{ irrational} \end{cases} \quad (5)$$

The Thomae function shown in Fig. 8b is discontinuous at each rational point, becomes zero, and is continuous at all irrational points.

Starting from the point of "subcluster percolation" $r \geq 2/3$, the distribution has an ultrametric structure typical of the statistics of extreme events. Analogous distributions arise in a number of other problems, for example, in the spectral statistics of the ensemble of exponentially weighed graphs [45–47], the profile of leaf growth with exponential cell division [48], the frequency of occurrence of certain subgraphs in the network of protein-protein interactions of *Drosophila* [49], and in a number of clinical data [50]. Without going into the mathematical discussions of similar distributions, let us note that all of them have characteristic features of the so-called Thomae function and may be regularized using the Dedekind η function near the real axis [47].

## CONCLUSIONS

In this study, we have considered the collapse of sufficiently long polymer chains with the simultaneous formation of crosslinks. Simulating the behavior of chains with different lengths at the same crosslinking protocol (time dependences of crosslink formation probability), we managed to distinguish different dependences of the characteristic subchain size on its length. For chains with length $N = 10^4$, we may distinguish two regimes: the regime of ideal subchain $R_g \sim s^{1/2}$ and the regime of random crosslink tree growth $R_g \sim s^{1/4}$, whereas for longer chains $N = 10^5$, the crumpled globule regime $R_g \sim s^{1/3}$ appears which is associated, on one hand, with the hierarchical structure of clusters imposed by crosslinks and, on the other hand, with the excluded volume of the polymer surrounding the subchain. An analogous effect is observed in the simulation of the ring polymer collapse: at a small polymer length, the presence of the surface prevents the development of a stable compact configuration.

Without setting a priori any properties of the real chromatin, the dependence of contact probability similar to the experimental dependence obtained for dense chromatin domains (TADs) in a certain scale range $P_c \sim s^{-3/4}$ was derived [34]. Using our method of the cluster analysis of pair contact maps, the simulation contact maps were annotated and it was found that clusterization makes itself more evident in globules obtained as a result of kinetic crosslinking (the kinetic freezing of contacts). Proceeding from this evidence, we may state that our results shed light on the hypothesis that local interactions (crosslinks) between remote regions of chromatin are one of the main factors controlling its three-dimensional structure and apparently are the main reason why the dependence of contact probability observed experimentally is typical of conformations with fractal dimension $D_f = 4$.

The evident observation is that, because of the discrete nature of the polymer chain and contacts in the clusters of individual maps, density (3) in the subdivision of contact maps into clusters with gradually decreasing density may assume only discrete rational values. Because of number-theoretic reasons not related to the physics of map construction, some density values are encountered in the map more frequently than others—these values set transition $r_{cr}^{(i)}$ values; once they are achieved, a fairly large number of sub-clusters are missing and the map division changes substantially. Indeed, the reduced ratio of a smaller natural number $a$ to a larger one $b$ equal to $\frac{a}{b} = \frac{p}{q}$, $\gcd(p,q) = 1$ is encountered on segment $a, b \in [1, L]$ $\left[\frac{L}{q}\right]$ times, where [...] denotes the integer part of number. Thus, the values of density with a smaller denominator (after reduction) will be encountered more often. It should be kept in mind that distribution of the number of contacts in subclusters is nonuniform; however, the degree of "degeneracy" (i.e., frequency) determined by law $\frac{1}{q}$ will nevertheless remain. For sufficiently high densities $r_{\mathrm{cr}}$, it is reasonable to assume that the probability of formation of the dense subcluster with size $Z$ and a small number of "voids" has the Poisson distribution $e^{-cS}$, where $S = Z^2$ is the subcluster area, and $c \ll 1$ determines the average concentration of "voids" (below it will be taken to be constant for different densities of subclusters $r > r_{\mathrm{cr}}$). By introducing the designation $f = e^{-c} \approx 1 - c$ (at $0 < c \ll 1$), we will take into account that the number of contacts $np$ will be encountered in the subcluster of area $Z^2 = nq$ with probability $f^{nq}$. In the limit $f \to 1$, this gives the distribution $1/q$:

$$g\left(\frac{p}{q}\right) = \sum_{n=1}^{L^2/q} f^{nq} \sim \left(\log \frac{1}{f}\right)^{-1} \frac{1}{q} = \frac{1}{cq}. \quad (6)$$

Thus, the values of critical density $r_{\mathrm{cr}}$ determining a sharp reduction in the average size of the clusters and their number correspond to the main sequence of peaks $\frac{k}{k+1}$, $k = 2, 3, 4, \ldots$ in distributions of the Thomae function type. The height of peaks characterizes the number of subclusters with a given density in the chain (Fig. 8). For density values corresponding to intermediate peaks, changes in $\langle L_r \rangle$ and $N_r$ are less pronounced, because the corresponding subclusters are encountered more rarely in individual contact maps.

The cluster analysis of intramolecular contact maps makes it possible to the conjecture that there is a set of characteristic discrete hierarchical levels in polymer packing which is a manifestation of the number-theoretic origin of rare-event statistics and is apparently inherent to any subdivisions of individual maps of intra- and interchromosomal contacts.

## ACKNOWLEDGMENTS

We are grateful to V.A. Avetisov, A. Gorskii, and V.A. Ivanov for valuable discussion.

This work was supported in part by the Federal Agency for Scientific Organizations of Russia within the scope of state task (Theme 0082-2014-0001, no. AAAA-A17-117040610310-6) and by the Russian Science Foundation (project no. 14-13-00745) using computational resources of the supercomputer complex of Moscow State University [51] (A. Astakhov); in part by the Russian Foundation for Basic Research, project no. 16-02-00252A (S. Nechaev); and in part by the Foundation for the Advancement of Theoretical Physics and Mathematics "BASIS" grant no. 17-12-278 (K. Polovnikov).

## REFERENCES

1. M. V. Vol'kenshtein, *Configurational Statistics of Polymer Chains* (AN SSSR, Moscow, 1959) [in Russian].
2. T. M. Birshtein and O. B. Ptitsyn, *Conformations of Macromolecules* (Nauka, Leningrad, 1964) [in Russian].
3. I. M. Lifshits, Sov. Phys. JETP **28**, 1280 (1969).
4. E. Zhulina, O. Borisov, and T. Birshtein, J. Phys. II **2**, 63 (1992).
5. A. Polotsky, M. Charlaganov, F. Leermakers, M. Daoud, O. Borisov, and T. Birshtein, Macromolecules **42**, 5360 (2009).
6. B. Mandelbrot, *The Fractal Geometry of Nature* (W. H. Freeman, San Francisco, 1982).
7. A. Yu. Grosberg, S. K. Nechaev, and E. I. Shakhnovich, J. Phys. (Paris) **49**, 2095 (1988).
8. A. Grosberg, Y. Rabin, S. Havlin, and A. Neer, Europhys. Lett. **23**, 373 (1993).
9. E. Lieberman-Aiden, N. L. van Berkum, L. Williams, M. Imakaev, T. Ragoczy, A. Telling, I. Amit, B. R. Lajoie, P. J. Sabo, M. O. Dorschner, R. Sandstrom, B. Bernstein, M. A. Bender, M. Groudine, A. Gnirke, J. Stamatoyannopoulos, L. A. Mirny, E. S. Lander, and J. Dekker, Science **326**, 289 (2009).
10. J. Dekker, K. Rippe, M. Dekker, and N. Kleckner, Science **295**, 1306 (2002).
11. K. Polovnikov, S. Nechaev, and M. V. Tamm, Soft Matter **14**, 6561 (2018).
12. K. Polovnikov, M. Gherardi, M. Cosentino-Lagomarsino, and M.V. Tamm, Phys. Rev. Lett. **120**, 088101 (2018).
13. M. Imakaev, S. Nechaev, K. Tchourine, and L. Mirny, Soft Matter **11**, 665 (2015).
14. R. D. Schram, G. T. Barkema, and H. Schiessel, J. Chem. Phys. **138**, 224901 (2013).
15. A. M. Astakhov, V. A. Ivanov, and V. V. Vasilevskaya, Dokl. Phys. Chem. **472**, 6 (2017).
16. R. K. Sachs, G. van der Engh, B. Trask, H. Yokota, and J. E. Hearst, Proc. Natl. Acad. Sci. U. S. A. **92**, 2710 (1995).
17. C. Münkel and J. Langowski, Phys. Rev. E: Stat. Phys., Plasmas, Fluids, Relat. Interdiscip. Top. **57**, 5888 (1998).
18. J. Ostashevsky, Mol. Biol. Cell **9**, 3031 (1998).
19. J. Mateos-Langerak, M. Bohn, W. de Leeuw, O. Giromus, E. M. M. Manders, P. J. Verschure, M. H. G. Indemans, H. J. Gierman, D. W. Heerman, R. van Driel, and

S. Goetze, Proc. Natl. Acad. Sci. U. S. A. **106**, 3812 (2009).

20. B. V. S. Iyer and G. Arya, Phys. Rev. E: Stat., Nonlinear, Soft Matter Phys. **86**, 011911 (2012).
21. M. Barbieri, M. Chotalia, J. Fraser, L.-M. Lavitas, J. Dostie, A. Pombo, and M. Nicodemi, Proc. Natl. Acad. Sci. U. S. A. **109**, 16173 (2012).
22. C. C. Fritsch and J. Langowski, Chromosome Res. **19**, 63 (2011).
23. A. Rosa and R. Everaers, PLoS Comput. Biol. **4** (2008).
24. L. A. Mirny, Cromosome Res. **19**, 37 (2011).
25. J. D. Halverson, J. Smrek, K. Kremer, and A. Yu. Grosberg, Rep. Prog. Phys. **77**, 022601 (2014).
26. A. Yu. Grosberg, Soft Matter **10**, 560 (2014).
27. A. Rosa and R. Everaers, Phys. Rev. Lett. **112**, 118302 (2014).
28. M. Tamm, L. Nazarov, A. Gavrilov, and A. Chertovich, Phys. Rev. Lett. **114**, 178102 (2015).
29. V. A. Avetisov, L. Nazarov, S. K. Nechaev, and M. V. Tamm, Soft Matter **11**, 1019 (2015).
30. J. R. Dixon, S. Selvaraj, F. Yue, A. Kim, Y. Li, Y. Shen, M. Hu, J. S. Liu, and B. Ren, Nature **485**, 7398 (2012).
31. E. P. Nora, J. Dekker, and E. Heard, BioEssays **35**, 9 (2013).
32. S. V. Ulianov, E. E. Khrameeva, A. A. Gavrilov, I. M. Flyamer, P. Kos, E. A. Mikhaleva, A. A. Penin, M. D. Logacheva, M. V. Imakaev, A. Chertovich, M. S. Gelfand, Y. Y. Shevelyov, and S. V. Razin, Genome Res. **26**, 1 (2016).
33. S. S. Rao, M. H. Huntley, N. C. Durand, E. K. Stamenova, I. D. Bochkov, J. T. Robinson, A. L. Sanborn, I. Machol, A. D. Omer, E. S. Lander, and E. L. Aiden, Cell **159**, 1665 (2014).
34. A. L. Sanborn, S. S. Rao, S. C. Huang, N. C. Durand, M. H. Huntley, A. I. Jewett, I. D. Bochkov, D. Chinnappan, A. Cutkosky, J. Li, K. P. Geeting, A. Gnirke, A. Melnikov, D. McKenna, E. K. Stamenova, E. S. Lander, and E. L. Aiden, Proc. Natl. Acad. Sci. U. S. A. **112**, E6456 (2015).
35. G. Fudenberg, M. Imakaev, C. Lu, A. Goloborodko, N. Abdennur, and L. A. Mirny, Cell Rep. **15**, 2038 (2016).
36. C. A. Brackley, J. Johnson, D. Michieletto, A. N. Morozov, M. Nicodemi, P. R. Cook, and D. Marenduzzo, Phys. Rev. Lett. **119**, 138101 (2017).
37. O. Shukron and D. Holcman, PLoS Comput. Biol. **13**, E1005469 (2017).
38. V. F. Scolari, G. Mercy, R. Koszul, A. Lesne, and J. Mozziconacci, Phys. Rev. Lett. **121**, 057801 (2018).
39. G. Bunin and M. Kardar, Phys. Rev. Lett. **115** (8), 088303 (2015).
40. S. Plimpton, J. Comput. Phys. **117**, 1 (1995).
41. A. Khokhlov and S. Nechaev, Phys. Lett. A **112**, 156 (1985).
42. T. Ge, S. Panyukov, and M. Rubinstein, Macromolecules **49** (2), 708 (2016).
43. E. Helfand and D. S. Pearson, J. Chem. Phys. **79**, 2054 (1983).
44. M. E. Cates and J. M. Deutsch, J. Phys. (Paris) **47**, 2121 (1986).
45. V. Avetisov, P. Krapivsky, and S. Nechaev, J. Phys. A: Math. Theor. **49**, 035101 (2016).
46. V. Kovaleva, Yu. Maximov, S. Nechaev, and O. Valba, J. Stat. Mech.: Theory Exp. **2017**, 073402 (2017).
47. S. Nechaev and K. Polovnikov, Phys.-Usp. **61**, 99 (2018).
48. S. Nechaev and K. Polovnikov, Soft Matter **13**, 1420 (2017).
49. M. Middendorf, E. Ziv, and C. H. Wiggins, Proc. Natl. Acad. Sci. U. S. A. **102**, 3192 (2005).
50. V. Trifonov, L. Pasqualucci, R. Dalla-Favera, and R. Rabadan, Sci. Rep. **1**, 191 (2011).
51. V. Sadovnichy, A. Tikhonravov, Vl. Voevodin, and V. Opanasenko, *"Lomonosov": Supercomputing at Moscow State University, Contemporary High Performance Computing: From Petascale toward Exascale,* Ed. by J. S. Vetter (Chapman and Hall/CRC Computational Science, Boca Raton, 2013), pp. 283−307.

*Translated by T. Soboleva*